\documentclass[aps,preprint,amsmath,amssymb,floatfix]{revtex4-2}
\usepackage{CJK}
\usepackage{graphicx}
\usepackage{dcolumn}
\usepackage{bm}
\usepackage{xcolor}
\usepackage[T1]{fontenc}
\usepackage{lmodern}
\usepackage[colorlinks=true,linkcolor=blue,urlcolor=blue,citecolor=blue]{hyperref}

\begin{document}

\title{Relaxation and Recovery in Hydrogel Friction on Smooth Surfaces
}


\author{Brady Wu}%
\author{Joshua M\'endez Harper}%
\author{Justin C. Burton}%
\email[email: ]{justin.c.burton@emory.edu}
\affiliation{Department of Physics, Emory University}
\date{\today}

\begin{abstract}
\textbf{Background} Hydrogels are crosslinked polymer networks that can absorb and retain a large fraction of liquid. Near a critical sliding velocity, hydrogels pressed against smooth surfaces exhibit time-dependent frictional behavior occurring over multiple timescales, yet the origin of these dynamics is unresolved. \textbf{Objective} Here, we characterize this time-dependent regime and show that it is consistent with two distinct molecular processes: sliding-induced relaxation and quiescent recovery. \textbf{Methods} Our experiments use a custom pin-on-disk tribometer to examine poly(acrylic acid) hydrogels on smooth poly(methyl methacrylate) surfaces over a variety of sliding conditions, from minutes to hours. \textbf{Results} We show that at a fixed sliding velocity, the friction coefficient decays exponentially and reaches a steady-state value. The time constant associated with this decay varies exponentially with the sliding velocity, and is sensitive to any precedent frictional shearing of the interface. This process is reversible; upon cessation of sliding, the friction coefficient recovers to its original state. We also show that the initial direction of shear can be imprinted as an observable ``memory", and is visible after 24 hrs of repeated frictional shearing. \textbf{Conclusions} We attribute this behavior to nanoscale extension and relaxation dynamics of the near-surface polymer network, leading to a model of frictional relaxation and recovery with two parallel timescales. 
\end{abstract}

\maketitle

\section{Introduction}

Hydrogel consists of a solvent-saturated, crosslinked polymer network that exhibits unique frictional properties due to its ambiguous nature. When swollen with liquid, osmotic pressure forces from the hydrophillic nature of the polymer network give rise to macroscale elasticity. However, hydrogels can be over 90\% liquid by weight, leading to a self-lubricating slippery surface \cite{Gong_2006,Bonyadi2020}. When exposed to a sliding interface, this dichotomy results in non-trivial and time-dependent frictional behavior that is important for numerous applications including biomaterials \cite{Moore_2016,Moore_2015,Moore_2014,Rennie_2005,Lee_2001,Dong_2006,Hamidi_2008,Larson_2016}, biomechanics \cite{Greene_2011,Lieleg_2011,Kisiday_2002,Barth_2016}, soft robotics \cite{Beebe_2000,Sidorenko_2007,Sun_2012,Bauer_2013,Keplinger_2013,Kim_2015,Yuk_2017}, and industrial water management \cite{Zhu2014,Sun_2012,Abidin2012,Chen2015,Tongwa2012,Guilherme_2015,Rudzinski_2002,Kim_2010}. The friction coefficient associated with a given interface generally depends on contact geometry \cite{Saintyves_2016,Pandey_2016,Urzay_2007,Skotheim_2005,Skotheim_2004,Yashima_2014}, sliding velocity \cite{Pitenis_2014,Dunn_2015,Kim_2018,Kim_2016}, liquid viscosity \cite{Cuccia_2020} and hydration \cite{Reale_2017,Moore_2017}, polymer density \cite{Cuccia_2020,Urue_a_2015,Kim_2010}, and any physicochemical absorption of polymers to the sliding layer \cite{Gong_1998,Gong_2006}.  

Although hydrogels exhibit interesting bulk rheologies, sliding interfaces expose the near-surface polymers in the network to shear forces that are orders of magnitude larger than those in the bulk. Such strong forces can alter the polymers' entropic configurations and preferentially stretch them in the sliding direction \cite{Kim_2016,Pitenis_2014,Dunn_2015,Reale_2017,Cuccia_2020}. This deformation can result in time-dependent frictional behavior during a single experiment at constant velocity \cite{Kim_2016,Kim_2018,Kim2020,Cuccia_2020}. Although the polymers near the interface are generally sparse, forming a crosslinked network with a ``mesh size'' \cite{degennes} of order 5-500 nm \cite{Scherer1994,Tsuji2018,Kim_2010,Cuccia_2020}, local entanglements and crosslinking constraints may increase the extension and relaxation rates. In Kim et al. \cite{Kim_2016,Kim_2018,Kim2020} and Cuccia et al. \cite{Cuccia_2020} time-dependent frictional behavior was observed over minutes or longer. Furthermore, Cuccia et al. \cite{Cuccia_2020} showed that for hydrogels on smooth surfaces, the friction coefficient evolves over multiple timescales and is not sensitive to liquid viscosity, suggesting that the frictional evolution is not due to re-hydration of the interface. The authors suggested that topological polymer entanglements may give rise to a free energy landscape with sufficient complexity to explain the timescales of frictional relaxation and recovery. 

Here, we show how the friction between a hydrogel and a smooth surface evolves under a broad set of sliding conditions by rapidly varying the velocity and pre-shearing of the interface. During some experiments, the hydrogel interface is allowed to rest in order to examine the recovery of friction after sliding. For polyacrylic acid (PAA) hydrogels on poly(methyl methacrylate) (PMMA) surfaces, we find that the friction coefficient, $\mu$, only displays time dependent behavior between $v\approx3-30$ mm/s, where $v$ is the sliding velocity. For a given sliding velocity in this regime, $\mu$ decays exponentially and approaches a steady-state value, independent of the sliding history of the hydrogel. However, the time constant and initial value of the friction do depend on the sliding history. In particular, the time constant decreases exponentially with sliding velocity. Similarly, the recovery of $\mu$ is exponential in the waiting time elapsed since the last experiment. We propose a simple, phenomenological model that captures the salient features of this behavior. The model consists of two parallel dynamical processes (relaxation and recovery), with timescales that can vary with the sliding velocity. Finally, and surprisingly, we show how the initial sliding direction can be imprinted as a ``memory'' into the time evolution of $\mu$, even after 24 hrs of continuous shearing. 

\section{Experimental Methods}

\begin{figure}[!]
  \includegraphics[width=3.4in]{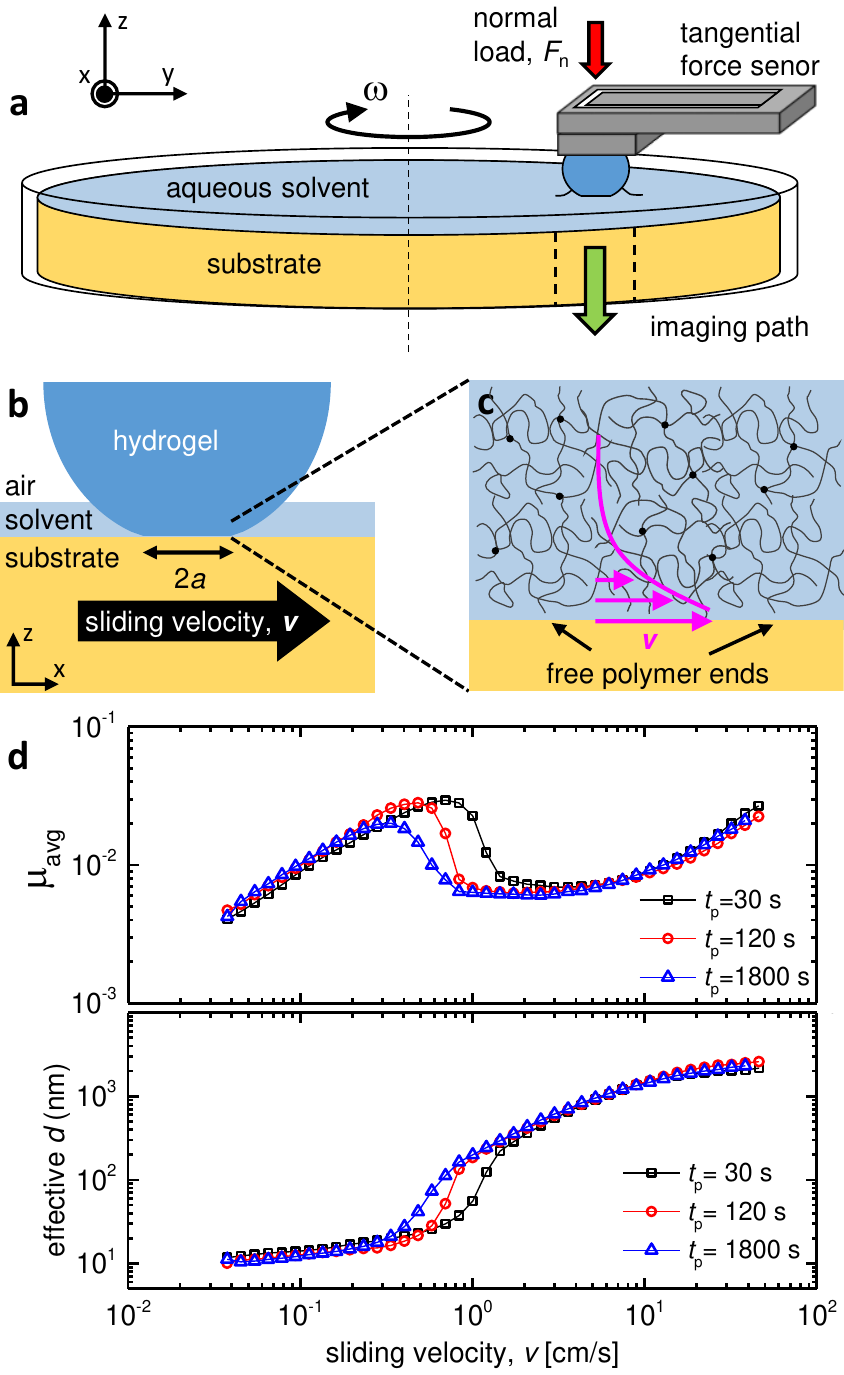}
\caption{(a) Diagram of the experimental setup showing the spherical sample pressed against the rotating surface. The hydrogel is sheared in the $x$-direction. (b) The characteristic diameter of the circular contact area is $2a$. (c) The polymer network adjacent to the surface experiences shear as fluid is dragged through it. (d) Plot of $\mu$ vs. $v$ for a PAA particle with $F_{n}$ = 0.2 N on an PMMA surface for three different values of the experimental running time, $t_{p}$, at each data point.  Also shown is the effective decay length of the velocity into the hydrogel, as calculated from Eq.\ \ref{fforce} using the viscosity of water ($8.9\times10^{-4}$ Pa.s). Error bars are not shown for clarity, but noise from the measurement apparatus is comparable to the point size.}
\label{fig:setup}     
\end{figure}

We performed experiments using the custom pin-on-disk, bi-directional tribometer, illustrated schematically in Fig.\ \ref{fig:setup}A. A similar device was used in Cuccia et al. \cite{Cuccia_2020}. A spherical sample -- PAA or agarose -- of radius $R\approx$ 7.5 mm was held stationary to the end of a low-force strain sensor (Strain Measurement Devices S256). This cantilevered spherical sample rested upon a horizontal PMMA substrate fixed to a circular, rotating frame.  Using atomic force microscopy, we determined that the root-mean-square roughness of the PMMA to be 3.8 nm over a 1 $\mu$m$^2$ area. The strain sensor measured the tangential force ($F_{f}$) on the sphere and was calibrated prior to use. A normal force $F_{n}=$ 200 mN was applied above the point of contact using a fixed amount of mass under the influence of gravity. A macroscopic layer of solvent over the substrate (thickness $\approx$ 3 mm) kept the contact hydrated during the experiment. We found that this amount of liquid led to repeatable friction measurements over multiple days, and also minimized any bulk drag from the fluid flow around the sphere.  The friction coefficient was calculated as $\mu=F_{{f}}/F_{n}$. All experiments were conducted at 22 $\pm$ 2$^\circ$C.

The experimental setup was automated using a LabVIEW program. To achieve a broad range rotational velocities at the point of contact between the spherical sample and the substrate, we drove the substrate with a stepper motor operating in a "microstepping" mode together a set of timing belts. This arrangement allowed us to investigate frictional behaviors at velocities ranging from 0.03 m/s and 60 m/s. The analog voltage at the output of the strain sensor was digitized using a USB data acquisition unit (NI USB-6008) at a rate of 24 kSamples per second. To reduce noise, 12,000 data points were averaged every 0.5 seconds. Thus, the friction coefficient was effectively measured two times per second.  Additionally, we were able to visually assess the point of contact through a milled hole in the structure underlying the spinning substrate. 

Individual experiments had durations ranging from 30 s $<t_p<$ 3600 s and the data from each trial was averaged to provide a single value of the friction coefficient $\mu_\text{avg}$. For some experiments (where indicated), the experiment was paused for 5 s, and then rotated in the opposite direction for the same time $t_{p}$ in order to average over both directions and calibrate the zero-point of the force sensor. At some velocities, the data displayed an exponential-like decay towards a steady-state value, leading to a large variation in the frictional coefficient. We particularly observe these effects within the transition regime when the friction decreased rapidly with sliding velocity.  We also observe that this decaying behavior is far more sensitive to changes in speed than changes in direction. 

We used two types of hydrogels for our experiments: poly(acrylic acid) (PAA) and agarose. Agarose samples were fabricated in the lab, and the PAA spherical particles were purchased from JRM Chemical (Cleveland, OH). The PAA particles were composed of approximately 70\% polyacrylic acid (PAA) and 30\% acrylamide monomer, as reported by the manufacturer.  Spherical agarose samples were made by mixing agarose (0.5-2.0 g) into 100 ml of distilled water and heating the solution to 60$^\circ$C. The solution was then pipetted into a spherical silicone mold and left aside at room temperature for 30 mins. The hydrogel was then immersed in water for at least 2 hrs. All chemicals and solvents used for fabrication were purchased from Millipore Sigma. We measured the ultimate swelling ratio of the gels by immersing them in water and letting them equilibriate. Dry commercial PAA particles, when immersed in water, would swell from an initial radius of $\approx 1$ mm to a final radius of $\approx 7.5$ mm.  Once swollen, the commercial PAA particles consisted of 99.2\% water. 

\section{Results}
\subsection{Regimes of frictional behavior}

First it is important to note the distinct regimes of frictional behavior we observe in hydrogels. At low velocities, the hydrogel is pressed firmly against the PMMA surface, making a circular contact area of radius $a\approx$ 3 mm (Fig.\ \ref{fig:setup}b). By varying the normal load and imaging the corresponding contact area using optical microscopy, we confirmed that the hydrogels obeyed a Hertzian relationship \cite{Cuccia_2020}. For a sphere pressed against a flat surface, the Hertz theory predicts that
\begin{equation}
a^3=\frac{3 F_{n} R}{4 E^*},
\end{equation}
where $E^*$ is the reduced average modulus for the sphere and the substrate:
\begin{equation}
\frac{1}{E^*}=\frac{1-\nu_{g}^2}{E_{g}}+\frac{1-\nu_{s}^2}{E_{s}}.
\label{modeq}
\end{equation}
Here $E$ and $\nu$ are the Young's modulus and Poisson's ratio for the gel (g) and substrate (s). The effective modulus for the PAA particles was measured as $E^*\approx$ 45 kPa, and for 2\% wt agarose it was measured as $E^*\approx$ 47 kPa \cite{Cuccia_2020}. 

As shown in Fig.\ \ref{fig:setup}c, when the hydrogel is adjacent to the surface, fluid is dragged through the porous polymer network. The characteristic decay length of the velocity is of order the mesh size \cite{Cuccia_2020}, so that the frictional force is proportional to the viscous shear stress at the interface:
\begin{equation}
\mu=\frac{A\eta}{F_n}\frac{v}{d}.
\label{fforce}
\end{equation}
Here $\eta$ is the dynamic viscosity of the solvent, $A=\pi a^2$ is the area of contact, and the velocity gradient has been replaced with $v/d$. At low velocities, this leads to a friction coefficient which increases with velocity due to viscous drag (Fig.\ \ref{fig:setup}d). The friction grows slower than linear with $v$ since $d$ can vary slightly with shear stress \cite{Cuccia_2020}. We can solve Eq.\ \ref{fforce} for $d$, and plot the result as a function of velocity, also shown in Fig.\ \ref{fig:setup}d. In the low-velocity regime, $d$ ranges from 10-20 nm, and the friction coefficient is independent of the amount of time spent measuring it at a given data point ($t_p$). Similarly, at high velocities, a bulk fluid layer of thickness $\approx$ 1 $\mu$m develops between the hydrogel and the PMMA surface. In this regime, the friction coefficient is determined by elastohydrodynamic lubrication theory (EHL), and is also independent of $t_p$. Both the low velocity and high velocity regimes are described in detail in Cuccia et al. \cite{Cuccia_2020}.

\begin{figure*}[!]
  \includegraphics[width=6.5in]{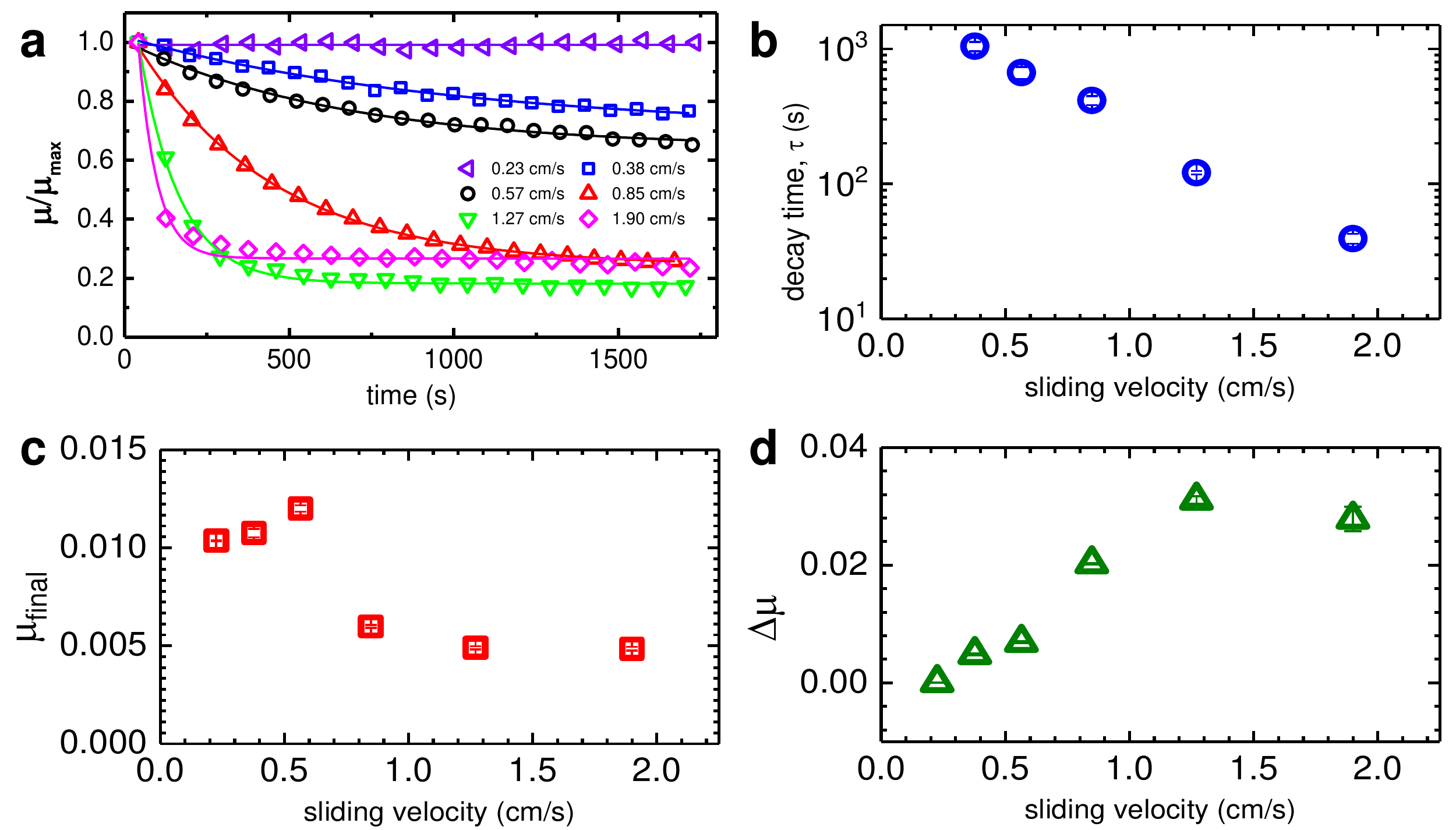}
\caption{(a) Normalized friction coefficient, $\mu/\mu_{max}$, versus time for a spherical PAA hydrogel on a PMMA surface at different sliding velocities. The hydrogel is allowed to ``rest" for 2 hrs prior to each experiment. (b-d) Velocity dependence of parameters obtained from fitting each data set to a single exponential form (Eq.\ \ref{exponential}). Both $\mu_f$ and $\Delta\mu$ approach constant values at high velocities, while $\tau$ decays exponentially.}
\label{fig:fresh_decay}       
\end{figure*}

\subsection{Relaxation of friction during sliding}

At intermediate velocities (0.3 cm/s $<v<$ 3.0 cm/s), we observe time-dependent dynamics. This can be seen in Fig.\ \ref{fig:setup}d, where the sharp drop in friction and transition to the EHL regime is sensitive to $t_p$. The variation of $\mu_\text{avg}$ with $t_p$ seen in Fig.\ \ref{fig:setup}d occurs because the friction decays exponentially with time during the course of an experiment at a single velocity \cite{Cuccia_2020}. At each new data point, $\mu_\text{avg}$ depends on the new velocity, but also on the history of shearing it has experienced. In order to tease apart this history dependence, we performed a series of experiments with a hydrogel particle that has been allowed to ``rest" for 2 hrs prior to sliding. Figure \ref{fig:fresh_decay}a shows the time evolution of $\mu$ after immediately shearing the hydrogel at different velocities for $t_p = 1800$ s. To compare the data, we normalize the data for each experiment by dividing by the maximum $\mu$ at $t=0$ s, denoted as $\mu_{max}$. For low velocities (0.23 cm/s), there is no variation in $\mu$ with time. However, at higher velocities, $\mu$ decreases and approaches a steady-state value. The rate of decay increases dramatically with velocity.

The data at each sliding velocity is fit remarkably well with a single exponential form,
\begin{equation}
\mu=\Delta\mu \times e^{-\frac{t}{\tau}}+\mu_f,
\label{exponential}
\end{equation}
as shown by the solid line in Fig.\ \ref{fig:fresh_decay}a. Here, $\mu_f$ is the final, steady state friction coefficient, $\Delta\mu=\mu_{max}-\mu_f$ is the change in friction, and $\tau$ is the decay time constant. Each of these fitting parameters varied significantly with the sliding velocity, as shown in Fig.\ \ref{fig:fresh_decay}b-d. Both $\Delta\mu$ and $\mu_f$ reached constant values at high velocities, 0.03 and 0.005, respectively. However, the decay time $\tau$ decreased rapidly in this range of sliding velocities, and follows an exponential dependence on $v$, as shown in Fig.\ \ref{fig:fresh_decay}b. 

This dependence suggests that the imposed shear may serve as an excitation mechanism to induce relaxation in the polymer network adjacent to the surface. A classical analogy is the $\alpha$-relaxation time for strong glass formers, which decreases exponentially with inverse temperature \cite{Angell1995}. Using Eq.\ \ref{exponential}, we can also compute the average friction coefficient over the time interval $t_p$, as was plotted in Fig.\ \ref{fig:setup}d:
\begin{equation}
\mu_\text{avg}=\mu_f+\Delta\mu\frac{\tau}{t_p}\left(1-e^{-t_p/\tau}\right).
\label{fric_avg}
\end{equation}
From this it can be seen why $\mu_\text{avg}$ drops so sharply Fig.\ \ref{fig:setup}d, since $\tau$ appears in the exponent, and also decreases exponentially with velocity.

\subsection{Recovery of friction during rest}

\begin{figure}[!]
  \includegraphics[width=3.4 in]{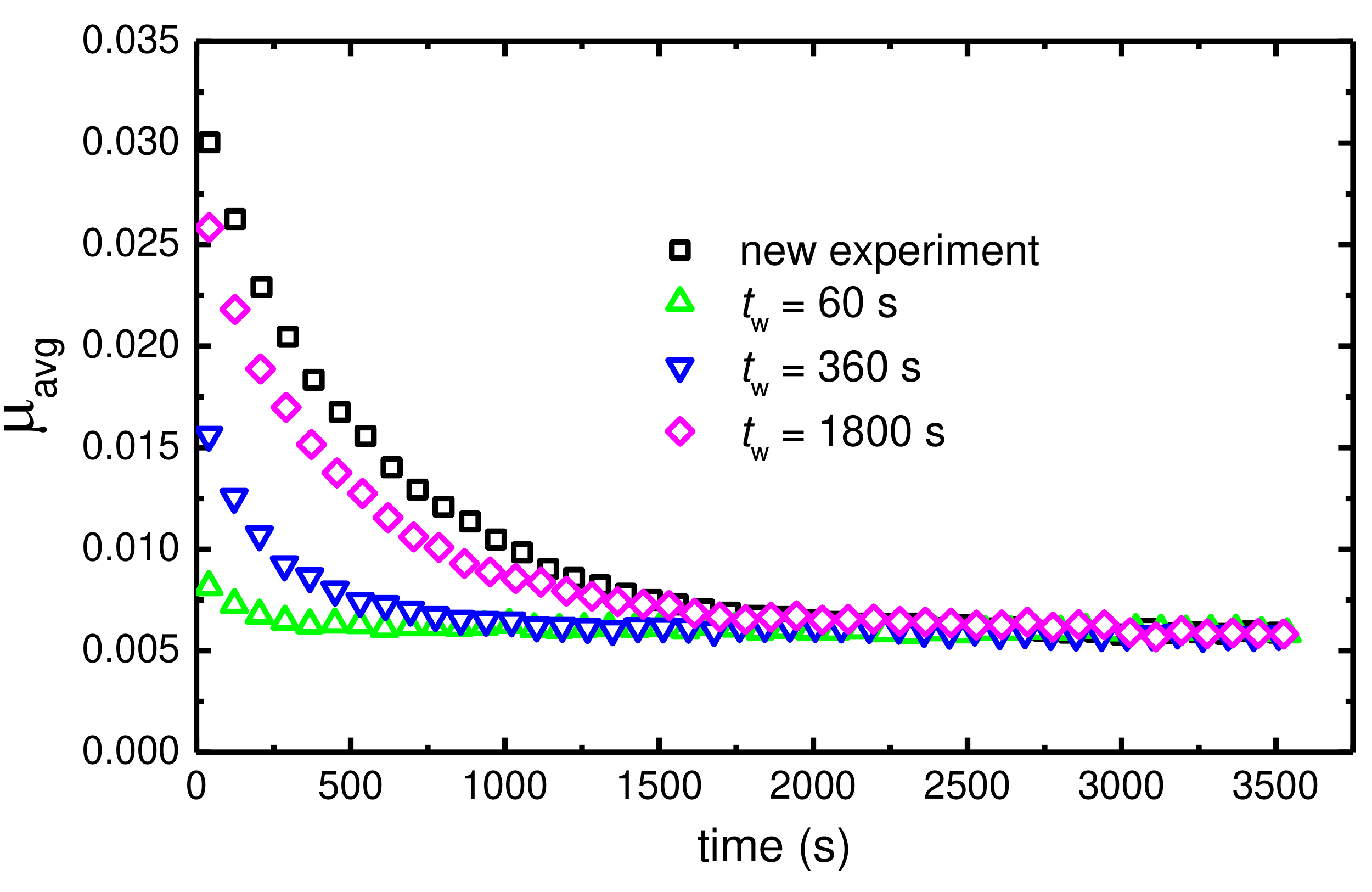}
\caption{The friction coefficient $\mu$ versus time for a spherical PAA hydrogel being shear at $v$ = 0.85 cm/s. ``New experiment" corresponds to a hydrogel that has rested for 2 hrs, whereas $t_w$ refers to the elapsed time after shearing that the hydrogel was allowed to wait prior to subsequent shearing. }
\label{fig:rec_vs_time}       
\end{figure}

In addition to exponential relaxation at sliding velocities in the intermediate regime, the friction coefficient also experiences a ``recovery" phase upon cessation of sliding. If the shear induced by sliding is able to preferentially align the near-surface polymers and modify the local entanglement structure of the network, then thermal fluctuations should facilitate a reversible transition to an equilibrium state in the absence of shear. Figure \ref{fig:rec_vs_time} shows a series of experiments where a fresh PAA hydrogel sphere is sheared for 1 hr, and then allowed to rest for a given waiting time $t_w$. At the end of each experiment, as in Fig.\ \ref{fig:fresh_decay}a, the hydrogel is able to reach its final friction coefficient ($\mu_f$). After waiting $t_w$ = 60 s, there is a small amount initial increase in the friction before it rapidly decays again. For larger values of $t_w$, both $\mu_{max}$ and $\tau$ increase so that the friction coefficient recovers more and it subsequently takes longer to fully decay again when sliding is initiated. Finally, after waiting 1800 s, the frictional evolution has almost fully recovered to that of a fully-rested hydrogel. 

\begin{figure}[!]
  \includegraphics[width=3.4in]{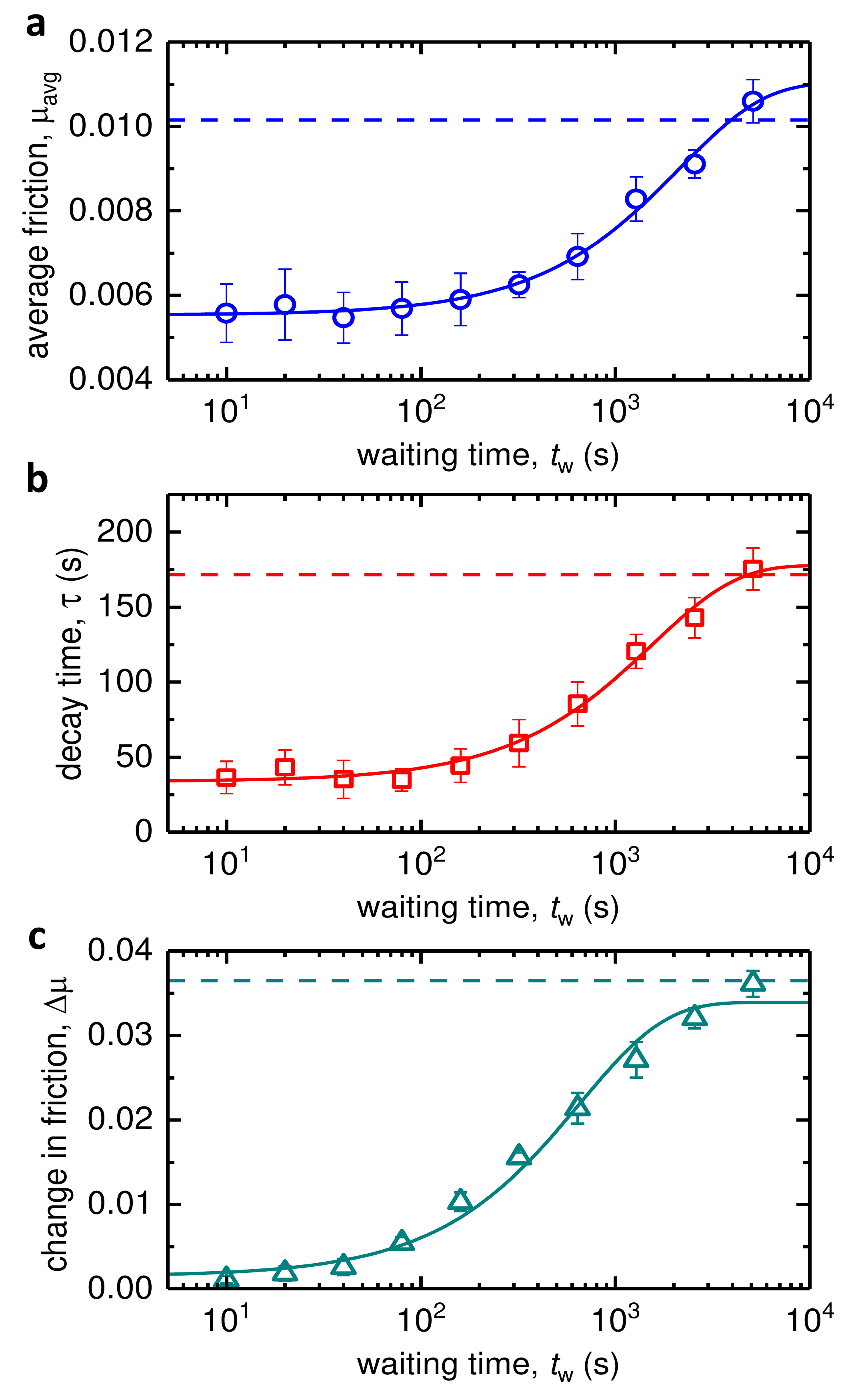}
\caption{(a) $\mu_\text{avg}$ versus $t_w$ for a PAA hydrogel sphere on PMMA. After initially shearing the hydrogel at $v$ 1.27 cm/s until it has reached a steady state value ($\mu_f$), the hydrogel is allowed to rest for a time $t_w$. Then data is acquired for $t_p$ = 600 s at $v$ = 1.27 cm/s for each data point. (b) The decay time, $\tau$, versus $t_w$, obtained from fitting each 600 s time series of $\mu$ to Eq.\ \ref{exponential}. (c) The change in friction coefficient, $\Delta \mu$, versus $t_w$, also obtained from fitting the data to Eq.\ \ref{exponential}. The solid lines represent exponential fits to the form $y=c_1(1-e^{-t_w/c_2})+c_3$, as described in the text. Error bars represent the standard deviations from three separate runs.}
\label{fig:recovery}       
\end{figure}

By fitting the time series of $\mu$ to the single exponential form (Eq.\ \ref{exponential}) for different values of $t_w$, we find that $\mu_\text{avg}$, $\tau$, and $\Delta\mu$ all recover to their initial values after waiting for a sufficiently long time. This is shown in Fig.\ \ref{fig:recovery}a-c. The dashed lines show the values associated with experiments from fresh, rested hydrogels. Importantly, even the decay time, $\tau$, seems to depend on the pre-shearing and $t_w$, changing by roughly a factor of 4 from short to long waiting times. This suggests that during recovery, the polymer network evolves over multiple timescales back to equilibrium. This is reasonable since structural relaxation in disordered systems typically contain a variety of timescales associated with relaxation to equilibrium. This comes from the ruggedness of the free energy landscape \cite{Berthier2011}. However, in systems with a broad range of timescales, relaxation is best described by a stretched exponential function rather than a single exponential. 

All of the data shown in Fig.\ \ref{fig:recovery}a-c is consistent with a generic, single exponential recovery: $y=c_1(1-e^{-t_w/c_2})+c_3$, where $c_1$, $c_2$, and $c_3$ are fitting parameters. These fits are shown by the solid lines. At the moment, we do not assign physical meaning to these parameters, but the origin of the fitting form will be discussed in Sec.\ \ref{sec:model}.

\subsection{Asymmetry of approach to $\mu_f$}

\begin{figure}[!]
  \includegraphics[width=3.4in]{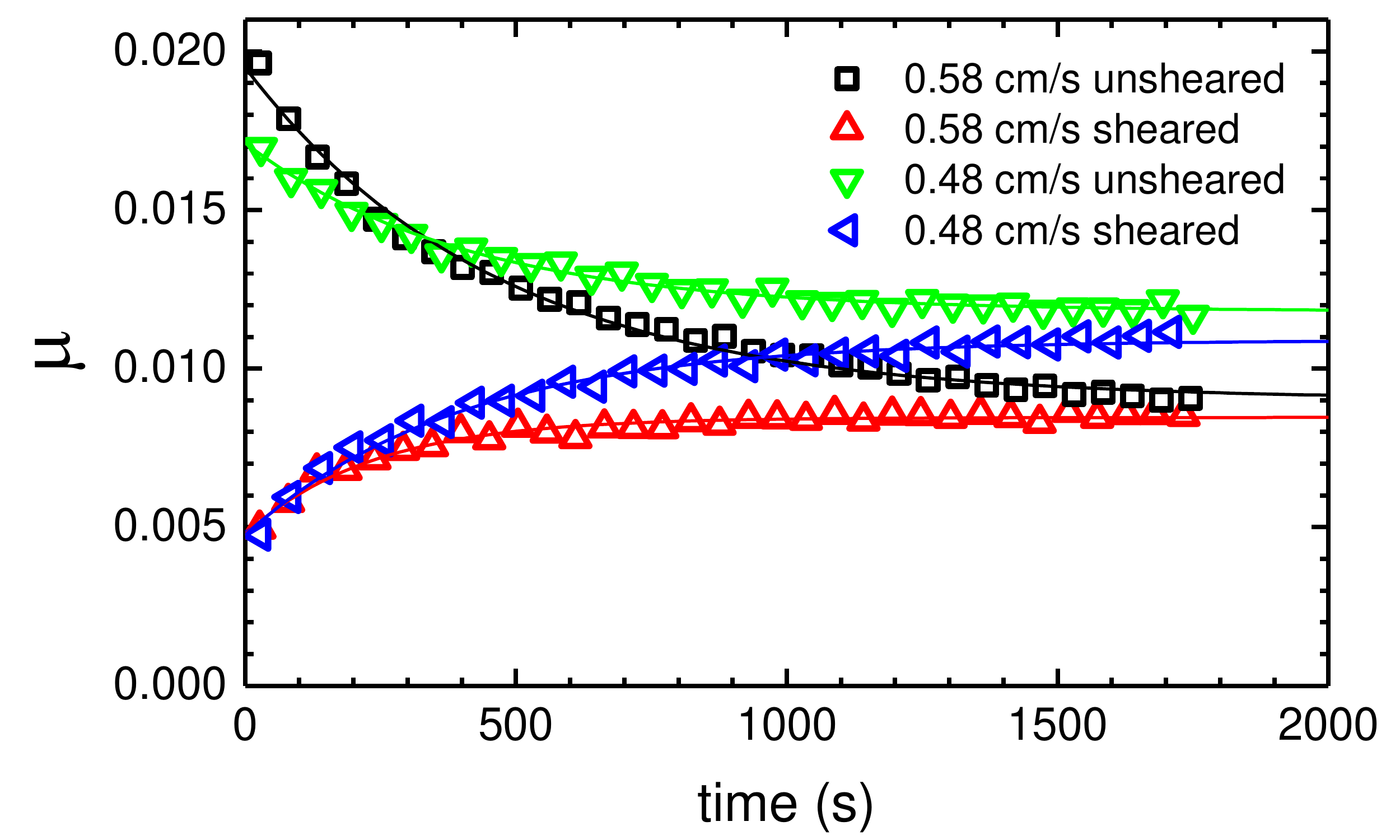}
\caption{Time series of the friction coefficient $\mu$ for PAA hydrogel on PMMA at with two distinct histories. The black squares and green downward triangles represent frictional relaxation starting from a fully rested gel at 0.58 cm/s and 0.48 cm/s, repsectively. The red upward triangles and blue sideways triangles represent relaxation after the hydrogel has been ``fully sheared" at $v$ = 3.0 cm/s for 1800 s. The approach to $\mu_text{final}$ is asymmetric, with different time constants. }
\label{fig:asymmetric}       
\end{figure}

The approach to a steady-state friction coefficient, $\mu_f$, depends on the shearing history of the hydrogel, and can be asymmetric. Figure \ref{fig:asymmetric} shows the decay in $\mu$ from two separate experiments at two different velocities. If the hydrogel is initially rested for 2 hrs before starting an experiment, $\mu_\text{avg}$ starts higher and decreases exponentially to $\mu_f$, as in Fig.\ \ref{fig:fresh_decay}. However, if the hydrogel is instead sheared at a high velocity, e.g. 3.0 cm/s, for 1800 s, the polymer network is presumably in the ``fully sheared" state, far from its equilibrium, rested value. In this regime, $\mu_f$ has plateaued with velocity (Fig.\ \ref{fig:fresh_decay}c). By immediately switching back to a lower velocity, the friction coefficient will then \emph{increase} with time, approach the same value of $\mu_f$. 

For both $v$ = 0.48 cm/s and $v$ = 0.58 cm/s in Fig.\ \ref{fig:asymmetric}, the initial value of $\mu$ and the decay constant $\tau$ is different, but the data for a given velocity tends to the same asymptotic value at long times. This behavior is reminiscent of the Kovacs effect in glassy, disordered materials \cite{Kovacs1963,Banik2018}. The effect is nonlinear in nature since perturbations away from equilibrium follow different relaxation mechanisms. Nevertheless, the fact that $\mu_f$ remains the same regardless of the shearing history suggests that the asymptotic sheared state of the system only depends on the sliding velocity.

\subsection{Model for frictional relaxation and recovery}

\label{sec:model}

The various timescales associated  with relaxation and recovery can at first seem complex. However, a simple, phenomenological model with two parallel timescales can capture the qualitative and some of the quantitative features of nearly all data for PAA hydrogel friction in the intermediate regime. We start by introducing a state parameter, $\gamma$, which has a range between 0 and 1. This parameterizes the sheared state of the hydrogel network at a given moment. In the fully ``rested" state with no shearing, $\gamma = 1$. For example, this characterizes the state of the polymer network in the low velocity regime, where fluid is dragged through the undeformed polymer network, and $\mu$ is determined by Eq.\ \ref{fforce}.

In order to be as quantitative as possible, we take into account that $\mu_\text{avg}$ is not perfectly linear in $v$ at low velocities (Fig.\ \ref{fig:setup}d). Thus, we use a power-law relationship to described $\mu$ in this regime: 
\begin{equation}
    \mu= K_1 v^{\alpha},
    \label{mu1}
\end{equation}
where $K_1$ is a fitting constant and $\alpha$ an exponent close to 1. In our model, we denote the friction coefficient of the hydrogel when $\gamma = 1$ as $\mu_{1}=K v^{\alpha}$. On the other hand, $\gamma = 0$ parameterizes the state where the hydrogel is fully sheared. We define a constant $\mu_0$ to be the lowest friction coefficient that the hydrogel can attain in the time dependent regime. For example, $\mu_f\approx0.005$ at $v=1.9$ cm/s in Fig.\ \ref{fig:fresh_decay}c. For simplicity, we assume that $\gamma$ varies linearly between $\mu_1$ and $\mu_0$, or 
\begin{equation}
    \gamma = \frac{\mu-\mu_0}{\mu_1-\mu_0}.
    \label{gamma}
\end{equation}

We now consider the relaxation and recovery processes separately. First, when perturbed by shear, thermal fluctuations will drive the polymers near the surface back to equilibrium ($\gamma=1$). As shown in Fig.\ \ref{fig:recovery}a, this process is well-captured by an exponential fit with a single time constant of order 1000-3000 s. We define this recovery time constant as $\tau_1$ as it refers to recovery to the $\gamma=1$ state. The evolution of $\mu$ is then given by:  
\begin{equation}
    \frac{d \mu}{dt}=\frac{\mu_1-\mu}{\tau_1},
    \label{recovery_eff}
\end{equation}
We note that the recovery time $\tau_1$ should be independent of sliding velocity, although it may depend on the fluid properties and temperature.

\begin{figure*}[!]
  \includegraphics[width=6.5in]{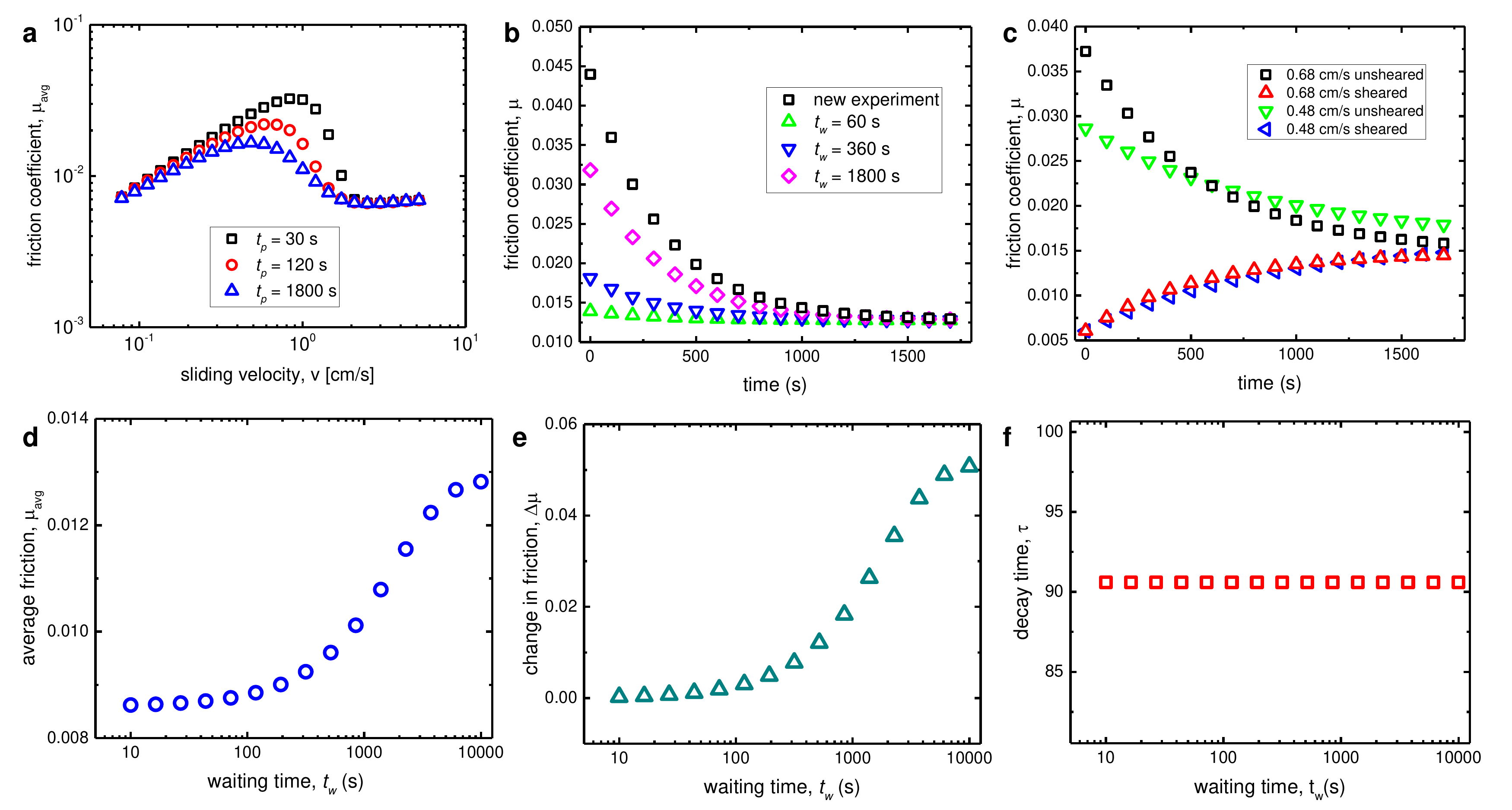}
\caption{Results from the analytic model (Eq.\ \ref{final_sol}) for different experimental protocols with $\tau_1$ = 1910 s, $K_0$ = 8887 s, $v_c$ = 0.276 cm/s, $\mu_0$ = 0.006, $\alpha$ = 3/4, and $K_1$ = 0.050 (s/cm)$^{3/4}$. (a) Starting from an initial, unsheared state ($\gamma=1$), $\mu_\text{avg}$ is computed over a time $t_p$. The value of $\mu$ at the end of each period is used as the intial condition for the next sliding velocity. (b) Recovery of friction with $v=0.85$ cm/s. After an initial evolution for 1800 s, the $\mu$ evolves with $v=0$ cm/s for a time $t_w$. Then sliding proceeds at the same velocity. The value of $\mu$ after the recovery period is used as the initial condition for each time series. (c) The green downward triangles and black squares show the evolution of $\mu$ at the given sliding velocities with no pre-shearing. After evolving $\mu$ at $v=3.0$ cm/s for 1800 s, $\mu$ is then allowed to evolve and approach the same steady-state value, demonstrating an asymmetry of approach. (d-e) $\mu_\text{avg}$, $\Delta\mu$, and $\tau$ plotted as a function of waiting time $t_w$. Starting from an unsheared state, $\mu$ evolves at $v=1.27$ cm/s for 600 s, then is allowed to recover for $t_w$. The value of $\mu$ at the end of each times series is used as the initial condition for the next time series.}
\label{fig:simulation}       
\end{figure*}

In the absence of recovery processes, the application of a sliding shear will lead to relaxation of $\mu$ to the minimum possible value, $\mu_0$. The evolution of $\mu$ is then given by 
\begin{equation}
    \frac{d \mu}{dt}=\frac{\mu_0-\mu}{\tau_0},
    \label{relaxation_eff}
\end{equation}
where $\tau_0$ is the time constant associated with relaxation to the $\gamma=0$ state. As suggested by Fig.\ \ref{fig:fresh_decay}b, we assume $\tau_0$ depends on the sliding velocity $v$:
\begin{equation}
    \tau_0=K_0e^{-v/v_c},
    \label{tau0}
\end{equation}
where the fitting constants $K_0$ and $v_c$ can be determined by fitting the data in Fig.\ \ref{fig:fresh_decay}b. 

During an experiment, these two processes are occurring in parallel. The full evolution of $\mu(t)$ is given by combining Eqs.\ \ref{recovery_eff} and \ref{relaxation_eff}:
\begin{equation}
    \frac{d \mu}{dt}=\frac{\mu_1-\mu}{\tau_1}+\frac{\mu_0-\mu}{\tau_0},
    \label{combined_eq}
\end{equation}
The measured dynamics of $\mu$ for a given experiment will depend on its initial shear state $\gamma$ and the sliding velocity $v$.  Since $\mu_1$ and $\tau_0$ don't explicitly depend on time (only on $v$), we can solve the equation analytically with the initial condition $\mu(t=0)=\mu_i$:
\begin{equation}
    \mu=(\mu_i-\mu_f) e^{-t/\tau}+ \mu_f.
    \label{final_sol}
\end{equation}
The solution is a single exponential decay where
\begin{equation}
    \mu_f=\frac{\tau_0\mu_1+\tau_1\mu_0}{\tau_1+\tau_0},
    \label{mu_fin}
\end{equation}
\begin{equation}
    \tau=\frac{\tau_1\tau_0}{\tau_1+\tau_0}.
    \label{tau_f}
\end{equation}

With this analytic form, we can recreate much of the data from the experiments. The empirical components of the model are the velocity-dependent forms of $\mu_1$ and $\tau_0$ (Eqs.\ \ref{mu1} and \ref{tau0}). Figure \ref{fig:simulation} shows analytic predictions which quantitatively capture many of our results. Panels (a-c) correspond to Figs.\ \ref{fig:setup}d, \ref{fig:rec_vs_time}, and \ref{fig:asymmetric}. Panels (d-f) correspond to Fig.\ \ref{fig:recovery}.  By choosing $\mu(t=0)$ and the sliding velocity $v$, an analytic time series can be created over a time $t_p$. This data then serves at the starting condition for the next data point in a simulated experiment shown in Fig.\ \ref{fig:simulation}.

One of the most important insights from this model is that the effective decay time $\tau$ observed during an experiment depends on a parallel combination of two independent time constants (Eq.\ \ref{tau_f}). Thus, $\tau$ will be smaller than either $\tau_1$ or $\tau_0$. This explains why we observe a variety of timescales for frictional dynamics in our experiments which depend on sliding velocity. Since Fig.\ \ref{fig:fresh_decay}b is technically reporting $\tau$ and not $\tau_0$, we convert it to $\tau$ using Eq.\ \ref{tau_f} and then apply Eq.\ \ref{tau0} for curve-fitting for all the simulation results reported above.


An important limitation of the model is evident in Fig.\ \ref{fig:simulation}f, which shows $\tau$ versus $t_w$. Since $\tau$ only depends on sliding velocity in our model and not on $t_w$, it is constant throughout the modeled experiment. However, as Fig.\ \ref{fig:recovery}b shows, 35 s $\lesssim\tau\lesssim$ 175 s throughout the physical experiment, implying that $\tau_1$ may depend somewhat on the state of the system, $\gamma$. This is reasonable since the perturbations from equilibrium can be quite large, and there can simultaneously exist slow and fast timescales due to the large deformation of the polymer network.

\subsection{Agarose and Directional Memories}

\begin{figure}[!]
  \includegraphics[width=3.3in]{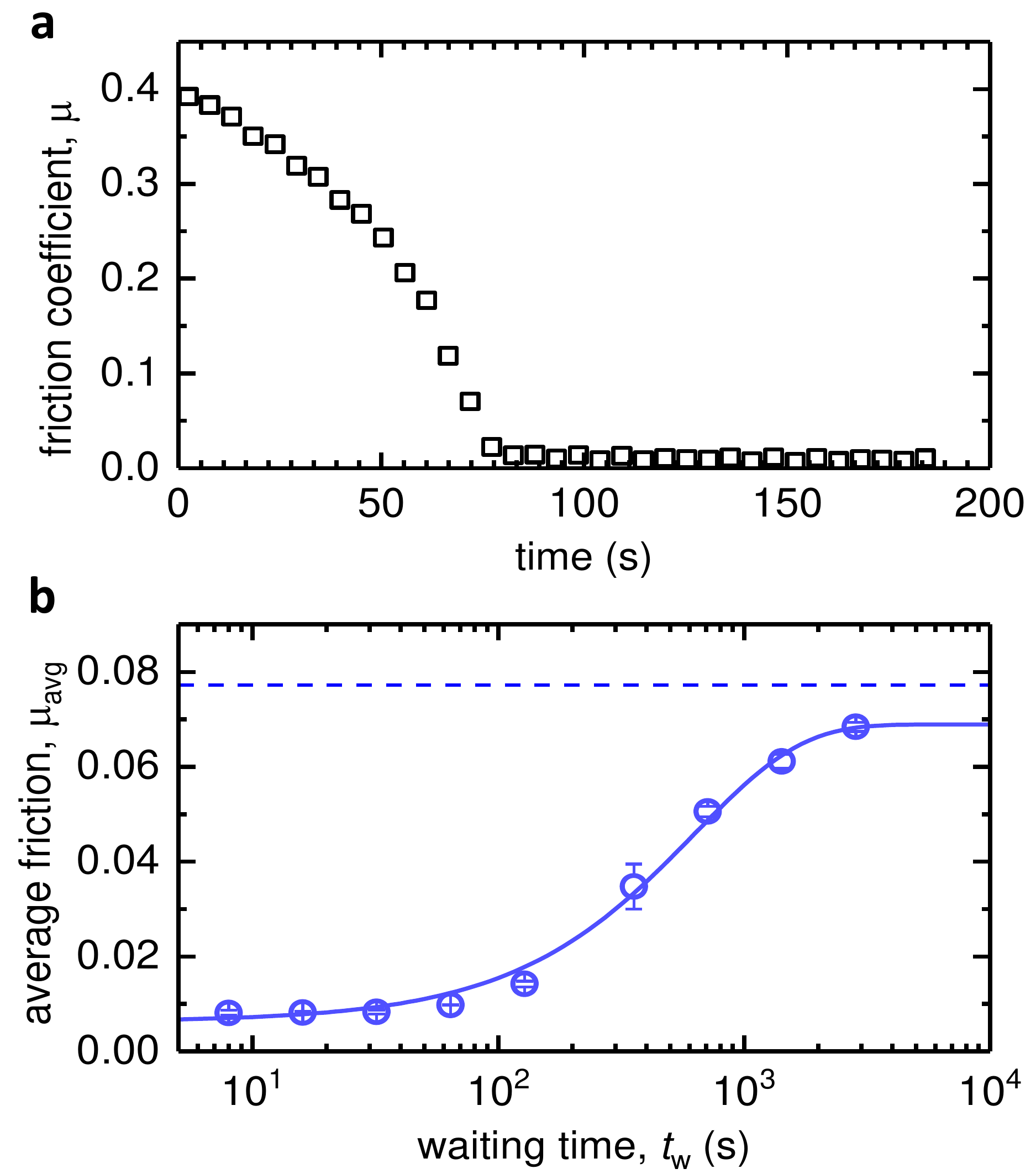}
\caption{(a) Friction coefficient of a 2\% agarose gel on PMMA versus time at $v=3.94 $ cm/s. The gel experiences no prior shearing, and the decay is not exponential. (b) The average friction coefficient of agarose gel, $\mu_\text{avg}$, at $v=3.94 cm/s$  versus waiting time. For each time series, $\mu_\text{avg}$ is computed for $t_p$ = 600 s, as in Fig.\ \ref{fig:recovery}. Error bars represent the standard deviations from three separate runs. }
\label{fig:agarose}      
\end{figure}

Although all of our experiments were performed with PAA hydrogel, we also tested agarose hydrogel spheres. In addition to acrylic acid and acrylamide hydrogels, Cuccia et al. \cite{Cuccia_2020} reported a similar rise in friction at low velocities for agarose spheres, followed by a sharp drop in friction in the time-dependent regime. However, the dynamics of $\mu$ versus $t$ for agarose hydrogels is quite distinct from PAA, and is not exponential. Figure \ref{fig:agarose}a shows that $\mu$ often starts at rather large values, near 0.4, then drops precipitously with an increasing rate until it reaches values near 0.01 after 100 s of sliding. For these experiments, $\Delta\mu$ can be quite large from the beginning to the end of a time series, more than 10 times the final value $\mu_f$. 

Nevertheless, the friction experiences similar recovery times as the PAA hydrogel, and is consistent with an exponential recovery process, as show in Fig.\ \ref{fig:agarose}a. This is consistent with hypothesized processes that determine relaxation and recovery. Agarose is a physical gel with transient crosslinks, suggesting that its near-surface network structure can be significantly altered due to shear, in contrast to the permanent crosslinks in PAA. We suspect that this may explain the sharp drop in $\mu$ in Fig.\ \ref{fig:agarose}a, and the non-exponential behavior. Rather than being perturbed from equilibrium, the local deformation and structural rearrangement is dramatic. However, if the recovery process is controlled by thermal fluctuations proportional to $k_\text{B} T$, then the recovery timescales can be similar to PAA. 

\begin{figure}[!]
  \includegraphics[width=3.3in]{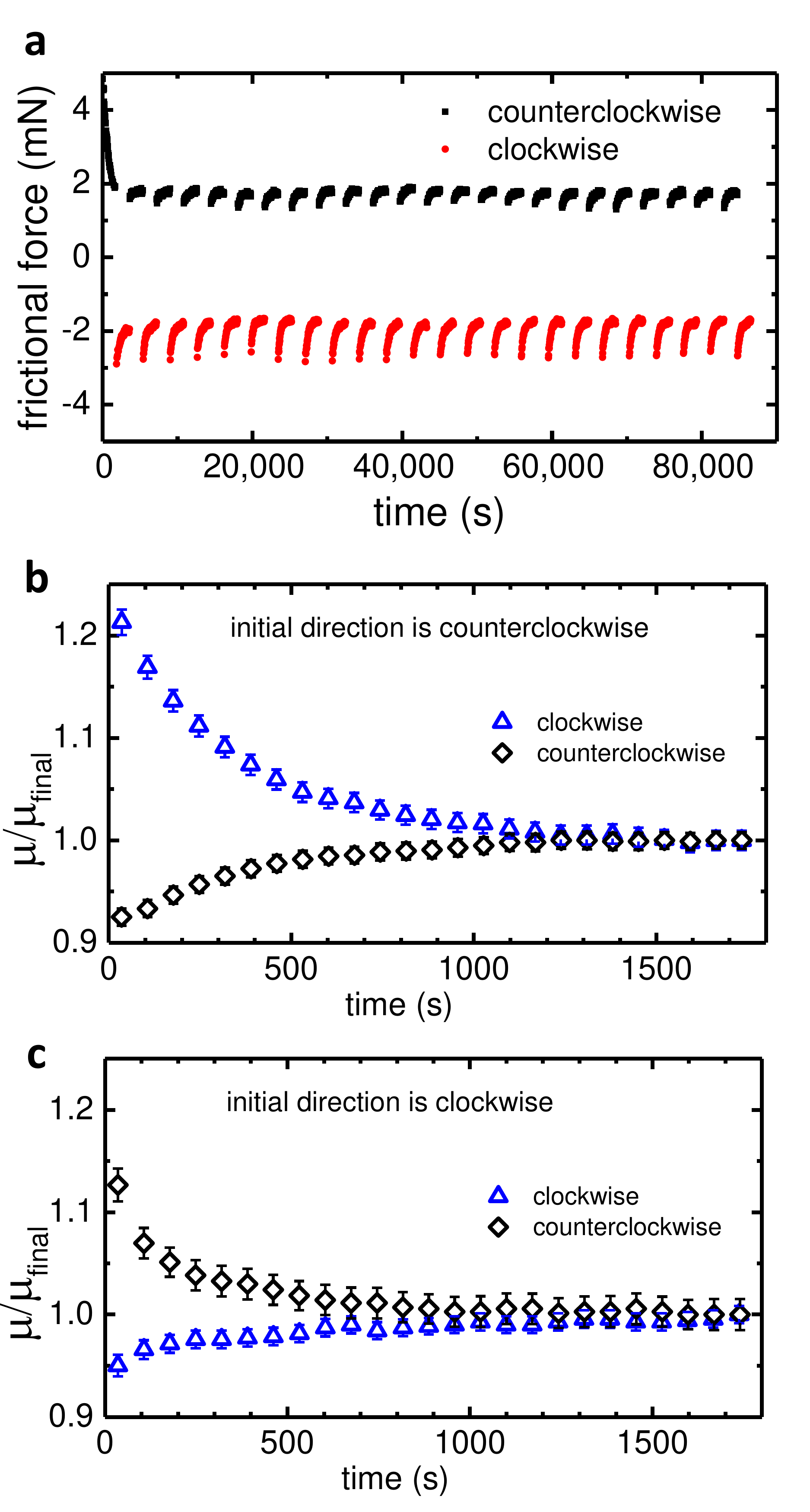}
\caption{Directional memories can be imprinted in the frictional evolution of PAA hydrogels. (a) The raw data of the frictional force from a PAA hydrogel sliding on a PMMA disk at $v=0.53$ cm/s. The inital sliding direction is CCW, and the direction is switched every 1800 s for a total period of 24 hrs. The directional asymmetry in the frictional evolution comes from the choice of the initial sliding direction. (b-c) Averaged data $\mu/\mu_f$ for CW and CCW directions showing that recovery always occurs in the initial sliding direction, and relaxation occurs in the opposing direction. }
\label{fig:memory}
\end{figure}

Finally, we explored the long-time behavior and directional dependence of PAA hydrogel friction. Starting with a fresh hydrogel sphere that did not experience pre-shearing, the frictional force was measured in both the counterclockwise (CCW) and clockwise (CW) direction, each for 1800 s at $v=0.53$ cm/s. After the initial decay of friction in the CCW direction, we observed relaxation in the CW direction, and recovery in the CCW direction, as shown in Fig.\ \ref{fig:memory}. Surprisingly, after 24 hrs, this asymmetry is still present, indicating that the initial direction of shear imprints a ``memory" that can be observed at long times.

To confirm that this is not an experimental artifact, we performed separate experiments with fresh hydrogels with different initial shear directions. Figure \ref{fig:memory}b-c shows that $\mu$ always experiences recovery when shearing in the initial direction, and relaxation in the opposing direction. However, $\mu_f$ was essentially the same in both directions, indicating that the final frictional state of the interface ($\gamma$ in our model) was independent of direction. We note that this asymmetry cannot be captured by our model in its current form since direction is not explicitly included. We suspect that this memory is imprinted by some semi-permanent orientational ordering at the hydrogel interface due to the initial sliding. It is unclear if this change would fully recovery given enough rest time, and warrants further investigation.

\section{Discussion and Summary}

The results presented here illustrate unique, time dependent frictional behavior between a hydrogel and a smooth, hard surface. In a range of intermediate sliding velocities, $\mu$ experiences a decay toward a new, velocity dependent value with a time constant that is itself velocity dependent. The frictional dynamics are reversible. The frictional state will recover given sufficient resting time, and there is no evidence of permanent wear. This behavior can be modeled quantitatively using two distinct time constants representing shear-induced relaxation and exponential recovery toward equilibrium. 

The analytic model proposed here is not too different from other ``rate and state" models used recently to describe time-dependent frictional behavior in hydrogels. However, the underlying mechanism leading to time-dependent behavior has not be fully resolved. Similar time-dependent behavior first observed in Kim et al. \cite{Kim_2016} was modeled using a frictional shear stress that followed two separate power-law behaviors according to a state variable that parameterized the rheological state of the interface \cite{Kim_2018,Kim2020}. A mechanism involving rehydration of the interface, characterized by a heterogenous distribution of bulk fluid pockets, was employed to explain the time and velocity dependence. However, Meier et al. \cite{Meier2019} and Simič et al. \cite{Simic2020} showed how the entropic freedom of near-surface polymers change alter $\mu$ by more than an order of magnitude. They also illustrate an exponential decay of $\mu$ whose time constant (while less than 1 minute) decreases with velocity. 

For the results reported here, the time constant associated with frictional relaxation varies by nearly 2 orders of magnitude over less than 1 order of magnitude in velocity. For two separate hydrogels with similar elastic modulii (PAA and agarose), the decay dynamics are different, one exponential (PAA), and one distinctly non-exponential (agarose). Moreover, the recovery of friction in these hydrogels, which has not been reported to our knowledge, occurs over remarkably similar timescales. This indicates that frictional sliding may activate the polymeric structure at the interface differently, yet their recovery pathways are determined by similar physics. For the time-dependent regime, we suspect that nanoscopic rehydration may facilitate a more ``brushy" interface associated with low friction \cite{Simic2020}, but ultimately the reversible, temporal evolution of $\mu$ is a feature of structural relaxation of the polymer network. Future studies involving direct observations of orientational molecular ordering at the interface may confirm this hypothesis. \\

\noindent \textbf{Acknowledgements} The authors would like to acknowledge support from NSF DMR Grant No. 14550869, and also from the Emory SURE summer program for undergraduate research.\\ 

\noindent \textbf{Compliance with Ethical Standards} The authors declare that they have no conflict of interest.

\bibliographystyle{spphys}       

\bibliography{hydrogel_relaxation}

\end{document}